\documentclass[a4paper,12pt]{article}
\usepackage{mathptm,times,graphicx}
\usepackage{amssymb,amsthm}
\usepackage{latexsym}
\usepackage{subfigure}

\newdimen\LENB \newdimen\LENW \newdimen\THI
\newdimen\LENWH \newdimen\LENTOT \newcount\N
\def\vbrknlnele#1#2#3{
  \LENB=#1pt \LENW=#2pt \THI=#3pt
  \LENWH=\LENW \divide\LENWH by 2
  \LENTOT=\LENB \advance\LENTOT by \LENW
  \vbox to \LENTOT{
    \vbox to \LENWH{}
    \nointerlineskip
    \vbox to \LENB{\hbox to \THI{\vrule width \THI height \LENB}}
    \nointerlineskip
    \vbox to \LENWH{}
  }}

\def\vbrknln#1{
  \N=#1
  \vcenter{
    \vbox{
      \loop\ifnum\N>0
        \vbox to 4pt{\vbrknlnele{2}{2}{0.1}}
        \nointerlineskip
        \advance\N by -1
      \repeat
  }}}

\def\hbrknlnele#1#2#3{
  \LENB=#1pt \LENW=#2pt \THI=#3pt
  \LENTOT=\LENB \advance\LENTOT by \LENW
  \vcenter{
    \vbox to \THI{
      \hbox to \LENTOT{
        \hfil
        \vrule width \LENB height \THI
        \hfil}
  }}}

\makeatletter

\usepackage{ulem}

\def\journal#1&#2,{\begingroup \let\journal=\dummyjournal
               \it #1\unskip~\bf\ignorespaces #2\rm,\endgroup}
\def\dummyjournal{\errmessage{Reference foul up: nested \journal macros}}

\def\eqref#1{(\ref{#1})}

\begin{document}
\title{Localized Solitons of 
a (2+1)-dimensional Nonlocal Nonlinear Schr\"odinger Equation}
\author{
Ken-ichi Maruno$^1\dagger$\\
$^1$
~Department of Mathematics, 
The University of Texas-Pan American,\\ 
Edinburg, TX 78539-2999\\
\quad \\
Yasuhiro Ohta$^{2}$\\
$^2$~Department of Mathematics, 
Kobe University, \\
Rokko, Kobe 657-8501, Japan
}
\date{\today}
\def\submitto#1{\vspace{28pt plus 10pt minus 18pt}
     \noindent{\small\rm To be submitted to : {\it #1}\par}}
\maketitle
\begin{abstract}
An integrable (2+1)-dimensional nonlocal 
nonlinear Schr\"odinger equation is discussed. 
The $N$-soliton solution is given by Gram type determinant. 
It is found that the localized 
N-soliton solution has interesting interaction behavior which shows
 change of 
amplitude of localized pulses after collisions. 
\par
\kern\bigskipamount\noindent
\end{abstract}

\kern-\bigskipamount


\section{Introduction}
\quad \\
The nonlinear Schr\"odinger (NLS) equation,
\begin{equation}
{\rm i}\psi_t=\psi_{xx}+\alpha|\psi|^2\psi\,, 
\label{contiNLS}
\end{equation}
is the most important soliton equation which 
is a widely used model for investigating the evolution of pulses in
optical fiber and of surface gravity waves with narrow-banded spectra in
fluid \cite{APT}. 
The study of vector and nonlocal analogues of the NLS 
equation has received considerable attention recently 
from both physical and mathematical 
points of view \cite{APT,pelinovsky,pelinovsky2,Kroli1,Kroli2,Deco}. 

In this Letter, we discuss a (2+1)-dimensional nonlocal 
nonlinear Schr\"odinger (2DNNLS) equation:
\begin{eqnarray}
{\rm i}u_t&=&u_{xx}+2u\int_{-\infty}^{\infty} |u|^2 dy\,,
\end{eqnarray}
where $u=u(x,y,t)$ is a complex function and $x,y,t$ are real. The Gram
type determinant solution is presented and localized 
soliton interactions are studied. 

\section{Determinant Solution}

Using the dependent variable transformation
\[
u(x,y,t)=\frac{g(x,y,t)}{f(x,t)}\,,\qquad
u^*(x,y,t)=\frac{g^*(x,y,t)}{f(x,t)}\,,\qquad 
\]
where $f$ is real and ${\,}^*$ is complex conjugate,
we have bilinear equations \cite{HirotaBook}
\begin{eqnarray}
&&(D_x^2-{\rm i}D_t)g\cdot f=0\,,\label{bilinear1}\\
&&(D_x^2+{\rm i}D_t)g^*\cdot f=0\,,\label{bilinear2}\\
&&D_x^2f\cdot f=2\int_{-\infty}^{\infty} gg^* dy\,.\label{bilinear3}
\end{eqnarray}
These bilinear equations have the following 
Gram determinant solution which is the $N$-soliton solution of the
2DNNLS equation:
\[
f={\left|\matrix{
\mathcal{A}_N &I_N
\cr\noalign{\vskip5pt}
-I_N & \mathcal{B}_N
}\right|}\,,
\]
\[
g={\left|\matrix{
\mathcal{A}_N &I_N & {\bf e}_N^T
\cr\noalign{\vskip5pt}
-I_N & \mathcal{B}_N & {\bf 0}^T
\cr\noalign{\vskip5pt}
{\bf 0} & -{\bf a}_N & 0
}\right|}\,,\quad 
g^*=-{\left|\matrix{
\mathcal{A}_N &I_N & {\bf 0}^T
\cr\noalign{\vskip5pt}
-I_N & \mathcal{B}_N & {{\bf a}^*}_N^T
\cr\noalign{\vskip5pt}
-{\bf e}_N^* & {\bf 0} & 0
}\right|}\,,
\]
where 
\[
\mathcal{A}_N=\left(\matrix{
\frac{{\rm e}^{\xi_1+\xi_1^{*}}}{p_1+p_1^*}
&\frac{{\rm e}^{\xi_1+\xi_2^{*}}}{p_1+p_2^*}&\cdots
&\frac{{\rm e}^{\xi_1+\xi_N^{*}}}{p_1+p_N^*}
\cr
\frac{{\rm e}^{\xi_2+\xi_1^{*}}}{p_2+p_1^*}
&\frac{{\rm e}^{\xi_2+\xi_2^{*}}}{p_2+p_2^*}&\cdots
&\frac{{\rm e}^{\xi_2+\xi_N^{*}}}{p_2+p_N^*}
\cr
\vdots&\vdots&\ddots&\vdots
\cr
\frac{{\rm e}^{\xi_N+\xi_1^{*}}}{p_N+p_1^*}
&\frac{{\rm e}^{\xi_N+\xi_2^{*}}}{p_N+p_2^*}&\cdots
&\frac{{\rm e}^{\xi_N+\xi_N^{*}}}{p_N+p_N^*}
\cr} \right)\,,
\]

\[
\mathcal{B}_N= 
\left(\matrix{
\frac{\int_{-\infty}^{\infty}a_1^*a_1dy}{p_1^*+p_1}
&\frac{\int_{-\infty}^{\infty}a_1^*a_2dy}{p_1^*+p_2}&\cdots
&\frac{\int_{-\infty}^{\infty}a_1^*a_Ndy}{p_1^*+p_N}
\cr
\frac{\int_{-\infty}^{\infty}a_2^*a_1dy}{p_2^*+p_1}
&\frac{\int_{-\infty}^{\infty}a_2^*a_2dy}{p_2^*+p_2}
&\cdots
&\frac{\int_{-\infty}^{\infty}a_2^*a_Ndy}{p_2^*+p_N}
\cr
\vdots&\vdots&\ddots&\vdots
\cr
\frac{\int_{-\infty}^{\infty}a_N^*a_1dy}{p_N^*+p_1}
&\frac{\int_{-\infty}^{\infty}a_N^*a_2dy}{p_N^*+p_2}&\cdots
&\frac{\int_{-\infty}^{\infty}a_N^*a_Ndy}{p_N^*+p_N}
}\right)\,,
\]
and $I_N$ is the $N\times N$ identity matrix, 
${\bf a}^T$ is the transpose of ${\bf a}$, 
\[
{\bf a_N}=(a_1,a_2,\cdots ,a_N)\,,\quad 
{\bf e_N}=(e^{\xi_1},e^{\xi_2},\cdots ,e^{\xi_N}) \,,\quad {\bf
0}=(0,0,\cdots , 0)\,,
\]
\[
\xi_i=p_ix-{\rm i}p_i^2t\,,\quad
\xi_i^*=p_i^*x+{\rm i}{p_i^*}^2t\,,
\qquad 1\le i\le N\,, 
\]
and 
$p_i$ is a complex wave number of $i$-th soliton and 
$a_i\equiv a_i(y)$ is a complex phase function of $i$-th soliton. 

Here, we show that eq.(\ref{bilinear3}) has the above Gram 
determinant solution.

Let us denote the $(i,j)$-cofactor of the matrix
\[
M=
{\left(\matrix{
\mathcal{A}_N &I_N
\cr\noalign{\vskip5pt}
-I_N & \mathcal{B}_N
}\right)}
\]
as $\Delta_{ij}$.
Then the $x$-derivative of $f=\det M$ is given by
\begin{eqnarray}
&&f_x=\sum_{i=1}^N\sum_{j=1}^N\Delta_{ij}\frac{\partial}{\partial x}
\frac{{\rm e}^{\xi_i+\xi_j^{*}}}{p_i+p_j^*}
=\sum_{i=1}^N\sum_{j=1}^N\Delta_{ij}{\rm e}^{\xi_i+\xi_j^{*}}
\nonumber\\
&&\quad =
{\left|\matrix{
\mathcal{A}_N &I_N & {\bf e}_N^T
\cr\noalign{\vskip5pt}
-I_N & \mathcal{B}_N & {\bf 0}^T
\cr\noalign{\vskip5pt}
 -{\bf e}_N^*&{\bf 0} & 0
}\right|}\,.
\end{eqnarray}
In the Gram determinant expression of $f$, dividing $i$-th row by
${\rm e}^{\xi_i}$ and multiplying $(N+i)$-th column by
${\rm e}^{\xi_i}$ for $i=1,\cdots,N$, and dividing $j$-th column by
${\rm e}^{\xi_j^{*}}$ and multiplying $(N+j)$-th row by
${\rm e}^{\xi_j^{*}}$ for $j=1,\cdots,N$, we obtain another
determinant expression of $f$,
\[
f=\det M'\,,
\]
where
\[
M'=
{\left(\matrix{
\mathcal{A'}_N &I_N
\cr\noalign{\vskip5pt}
-I_N & \mathcal{B'}_N
}\right)},
\]
\[
\mathcal{A'}_N=\left(\matrix{
\frac{1}{p_1+p_1^*}&\cdots
&\frac{1}{p_1+p_N^*}
\cr\noalign{\vskip5pt}
\vdots&\ddots &\vdots
\cr\noalign{\vskip5pt}
\frac{1}{p_N+p_1^*}&\cdots
&\frac{1}{p_N+p_N^*}
}\right)\,, 
\]
\[
\mathcal{B'}_N=\left(\matrix{
\displaystyle
\frac{{\rm e}^{\xi_1^{*}+\xi_1}\int_{-\infty}^{\infty}a_1^*a_1dy}{p_1^*+p_1}
&\cdots
&\displaystyle
\frac{{\rm e}^{\xi_1^{*}+\xi_N}\int_{-\infty}^{\infty}a_1^*a_Ndy}{p_1^*+p_N}
\cr
\vdots&\ddots&\vdots
\cr
\displaystyle
\frac{{\rm e}^{\xi_N^{*}+\xi_1}\int_{-\infty}^{\infty}a_N^*a_1dy}{p_N^*+p_1}
&\cdots
&\displaystyle
\frac{{\rm e}^{\xi_N^{*}+\xi_N}\int_{-\infty}^{\infty}a_N^*a_Ndy}{p_N^*+p_N}
}\right)\,.
\]

Thus the $x$-derivative of $f$ is also written as
\begin{eqnarray}
&&f_x=\sum_{i=1}^N\sum_{j=1}^N\Delta'_{N+i,N+j}\frac{\partial}{\partial x}
\frac{{\rm e}^{\xi_i^{*}+\xi_j}\int_{-\infty}^{\infty}a_i^*a_jdy}{p_i^*+p_j}
=\sum_{i=1}^N\sum_{j=1}^N\Delta'_{N+i,N+j}
{\rm e}^{\xi_i^{*}+\xi_j}\int_{-\infty}^{\infty}a_i^*a_jdy
\nonumber\\
&&\quad =\int_{-\infty}^{\infty}\sum_{i=1}^N\sum_{j=1}^N\Delta'_{N+i,N+j}
{\rm e}^{\xi_i^{*}+\xi_j}a_i^*a_jdy
\nonumber
\end{eqnarray}
where $\Delta'_{ij}$ is the $(i,j)$-cofactor of $M'$.
Therefore we have
\begin{eqnarray}
&&f_x=\int_{-\infty}^{\infty}
\left|\matrix{
\mathcal{A'}_N &I_N & {\bf 0}^T
\cr
-I_N & \mathcal{B'}_N & {{\bf {\tilde a}}_N}^{*T}
\cr
{\bf 0} & -{\bf {\tilde a}}_N & 0
}\right|dy
\nonumber\\
&&\quad =\int_{-\infty}^{\infty}
{\left|\matrix{
\mathcal{A}_N &I_N & {\bf 0}^T
\cr
-I_N & \mathcal{B}_N & {{\bf a}^*}_N^T
\cr
{\bf 0} & -{\bf a}_N & 0
}\right|}dy\,,
\nonumber
\end{eqnarray}
where
\[
 {\bf {\tilde a}}_N=(e^{\xi_1}a_1,\cdots,e^{\xi_N}a_N )\,.
\]
By differentiating the above $f_x$ by $x$, we get
\[
f_{xx}=\int_{-\infty}^{\infty}
\left|\matrix{
\mathcal{A}_N &I_N & {\bf e}_N^T & {\bf 0}^T
\cr\noalign{\vskip5pt}
-I_N & \mathcal{B}_N & {\bf 0}^T & {{\bf a}^*}_N^T
\cr\noalign{\vskip5pt}
 -{\bf e}_N^*&{\bf 0} & 0 & 0
\cr\noalign{\vskip5pt}
{\bf 0} & -{\bf a}_N & 0 & 0}
\right| dy.
\]
On the other hand, using the Jacobi formula for
determinant\cite{HirotaBook}, 
we have
\begin{eqnarray}
&&gg^*=
\left|\matrix{
\mathcal{A}_N &I_N
\cr\noalign{\vskip5pt}
-I_N & \mathcal{B}_N
}\right|\times
\left|\matrix{
\mathcal{A}_N &I_N & {\bf e}_N^T & {\bf 0}^T
\cr\noalign{\vskip5pt}
-I_N & \mathcal{B}_N & {\bf 0}^T & {{\bf a}^*}_N^T
\cr\noalign{\vskip5pt}
 -{\bf e}_N^*&{\bf 0} & 0 & 0
\cr\noalign{\vskip5pt}
{\bf 0} & -{\bf a}_N & 0 & 0}
\right|
\nonumber\\
&&\qquad -
\left|\matrix{
\mathcal{A}_N &I_N & {\bf e}_N^T
\cr\noalign{\vskip5pt}
-I_N & \mathcal{B}_N & {\bf 0}^T
\cr\noalign{\vskip5pt}
 -{\bf e}_N^*&{\bf 0} & 0
}\right|
\times
\left|\matrix{
\mathcal{A}_N &I_N & {\bf 0}^T
\cr
-I_N & \mathcal{B}_N & {{\bf a}^*}_N^T
\cr
{\bf 0} & -{\bf a}_N & 0
}\right|.
\nonumber
\end{eqnarray}
Here we note that the $y$-dependence in right-hand side appears only
in the last row and last column of the second determinant in each
term.  Thus we obtain
\[
\int_{-\infty}^{\infty}gg^*dy=ff_{xx}-f_xf_x\,,
\]
which is a bilinear equation (\ref{bilinear3}).

Since eqs.(\ref{bilinear1}) and (\ref{bilinear2}) are 
bilinear equations for the NLS equation and do not include $y$, 
we can prove in the same way in the NLS equation that 
the above Gram determinant solution satisfies the bilinear identities 
(\ref{bilinear1}) and (\ref{bilinear2}), i.e., we can 
show easily that 
the bilinear equations (\ref{bilinear1}) and (\ref{bilinear2}) 
are made from a pair of Jacobi identities, respectively. 

\section{Localized Solitons}

Using the above formula, we can make 1-soliton solution as follows: 
\begin{equation}
u=
\frac{g}{f}
=
\frac{a_1{\rm e}^{\xi_1}}{1+\frac{\int_{-\infty}^{\infty}
a_1^*a_1dy}{(p_1^*+p_1)^2}{\rm e}^{\xi_1+\xi_1^{*}}}\,,
\quad 
u^*=
\frac{g^*}{f}
=
\frac{a_1^*{\rm e}^{\xi_1^*}}{1+\frac{\int_{-\infty}^{\infty}
a_1^*a_1dy}{(p_1^*+p_1)^2}{\rm e}^{\xi_1+\xi_1^{*}}}\,,
\end{equation}
where
\[
f={\small \left|\matrix{
\displaystyle\frac{{\rm e}^{\xi_1+\xi_1^{*}}}{p_1+p_1^*}
&1
\cr\noalign{\vskip5pt}
-1&\displaystyle\frac{\int_{-\infty}^{\infty}
a_1^*a_1dy}{p_1^*+p_1}
\cr\noalign{\vskip5pt}
}\right|}=1+\frac{\int_{-\infty}^{\infty}
a_1^*a_1dy}{(p_1^*+p_1)^2}{\rm e}^{\xi_1+\xi_1^{*}}\,,
\]

\[ 
g={\small\left|\matrix{
\displaystyle\frac{{\rm e}^{\xi_1+\xi_1^{*}}}{p_1+p_1^*}
&1
&{\rm e}^{\xi_1}
\cr\noalign{\vskip5pt}
-1&\displaystyle
\frac{\int_{-\infty}^{\infty}a_1^*a_1dy}{p_1^*+p_1}
&0
\cr\noalign{\vskip5pt}
0&-a_1&0
}\right|}=a_1{\rm e}^{\xi_1}\,,
\]
\[
g^*=-
{\small \left|\matrix{
\displaystyle\frac{{\rm e}^{\xi_1+\xi_1^{*}}}{p_1+p_1^*}
&1&0
\cr\noalign{\vskip5pt}
-1&\displaystyle
\frac{\int_{-\infty}^{\infty}a_1^*a_1dy}{p_1^*+p_1}
&a_1^*
\cr\noalign{\vskip5pt}
-{\rm e}^{\xi_1^{*}}
&0&0
}\right|}=a_1^*{\rm e}^{\xi_1^*}\,.
\]

\begin{figure}[t!]
\centerline{
\includegraphics[scale=0.7]{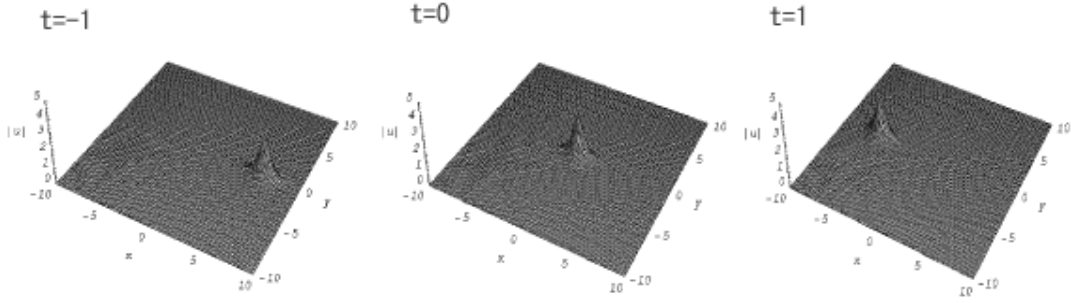}}
\caption{1-soliton solution. $\alpha_1=1+2i, p_1=2+3i, k=3,
 \eta_0=0$.}
\label{fig1}
\end{figure}

If we choose $a_1(y)=\alpha_1 {\rm sech}(k(y+\eta_0))$ where $\alpha_1$ is
a complex number and $k$ and $\eta_0$ are real numbers,
\[
u=
\frac{\alpha_1 {\rm sech}(k(y+\eta_0)){\rm e}^{\xi_1}}
{1+\frac{(2/k)|\alpha_1|^2}{(p_1^*+p_1)^2}{\rm e}^{\xi_1+\xi_1^{*}}}
=\frac{\alpha_1}{2\sqrt{A}}{\rm sech}(ky+\eta_0)
{\rm sech}\left(\frac{\xi_1+\xi_1^*}{2}+\frac{1}{2}\log A\right)
e^{\frac{\xi_1-\xi_1^*}{2}}\,,
\]
where $A=\frac{(2/k)|\alpha_1|^2}{(p_1^*+p_1)^2}\,$. 
In this case, we have a localized pulse as shown in figure \ref{fig1}.  

\begin{figure}[t!]
\centerline{
\includegraphics[scale=0.7]{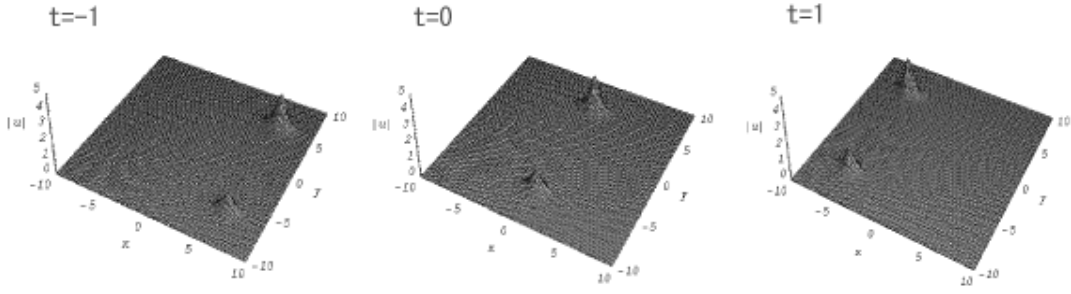}}
\caption{1-soliton solution. $\alpha_1=1+2i, \alpha_2=1/2+i, p_1=2+3i, k=3,
 \eta_1=-6, \eta_2=6$.}
\label{fig2}
\end{figure}

If we choose $a_1(y)=\alpha_1 {\rm sech}(k(y+\eta_1))
+\alpha_2 {\rm sech}(k(y+\eta_2))$ where $\alpha_1$ and $\alpha_2$ are
complex numbers and $k$, $\eta_1$ and $\eta_2$ are real numbers,
\begin{eqnarray}
&&
u=
\frac{(\alpha_1 {\rm sech}(k(y+\eta_1))
+\alpha_2 {\rm sech}(k(y+\eta_2))){\rm e}^{\xi_1}}
{1+\frac{(2/k)(|\alpha_1|^2+|\alpha_2|^2)+
4(\eta_1-\eta_2)(\alpha_1\alpha_2^*
+\alpha_1^*\alpha_2)/
({\rm e}^{k(\eta_1-\eta_2)}-{\rm e}^{-k(\eta_1-\eta_2)})}
{(p_1^*+p_1)^2}{\rm e}^{\xi_1+\xi_1^{*}}}\nonumber\\
&&\,=\frac{1}{2\sqrt{A}}
(\alpha_1 {\rm sech}(k(y+\eta_1))
+\alpha_2 {\rm sech}(k(y+\eta_2))\nonumber\\
&&\qquad \times 
{\rm sech}\left(\frac{\xi_1+\xi_1^*}{2}+\frac{1}{2}\log A\right)
e^{\frac{\xi_1-\xi_1^*}{2}}\,,\nonumber
\end{eqnarray}
where $A=\frac{(2/k)(|\alpha_1|^2+|\alpha_2|^2)+
4(\eta_1-\eta_2)(\alpha_1\alpha_2^*
+\alpha_1^*\alpha_2)/
({\rm e}^{k(\eta_1-\eta_2)}-{\rm e}^{-k(\eta_1-\eta_2)})}
{(p_1^*+p_1)^2}\,$. We see two localized pulses in figure \ref{fig2}. 
These two localized pulses travel parallel to the $x$-axis. 
With $a_1(y)=\sum_j^M \alpha_j {\rm sech}(k(y+\eta_j))$, 
we can see $M$-localized pulses travelling parallel to the $x$-axis. 
We call this $M$-localized pulse the 
$(1,M)$-localized pulse solution. In the general case of pulse solutions 
generated from the $N$-soliton formula, 
it is named by $(N,M)$-localized pulse solution. 

Next, we consider the case of $N=2$, i.e. 2-soliton solution. 
Using the determinant form of $N$-soliton solution, we have 
\begin{eqnarray*}
&&f={\small \left|\matrix{
\displaystyle\frac{{\rm e}^{\xi_1+\xi_1^{*}}}{p_1+p_1^*}
&\displaystyle\frac{{\rm e}^{\xi_1+\xi_2^{*}}}{p_1+p_2^*}
&1&0
\cr\noalign{\vskip5pt}
\displaystyle\frac{{\rm e}^{\xi_2+\xi_1^{*}}}{p_2+p_1^*}
&\displaystyle\frac{{\rm e}^{\xi_2+\xi_2^{*}}}{p_2+p_2^*}
&0&1
\cr
\noalign{\vskip5pt}
-1&0&\displaystyle\frac{\int_{-\infty}^{\infty}
a_1^*a_1dy}{p_1^*+p_1}
&\displaystyle\frac{\int_{-\infty}^{\infty}
a_1^*a_2dy}{p_1^*+p_2}
\cr\noalign{\vskip5pt}
0&-1&\displaystyle\frac{\int_{-\infty}^{\infty}
a_2^*a_1dy}{p_2^*+p_1}
&\displaystyle\frac{\int_{-\infty}^{\infty}
a_2^*a_2dy}{p_2^*+p_2}
}\right|}\nonumber\\
&&
\quad =1+
\frac{c_{11}}{p_1^*+p_1}{\rm e}^{\xi_1+\xi_1^{*}}
+\frac{c_{12}}{p_1^*+p_2}{\rm e}^{\xi_1^*+\xi_2}
+\frac{c_{21}}{p_2^*+p_1}{\rm e}^{\xi_1+\xi_2^*}
+\frac{c_{22}}{p_2^*+p_2}{\rm e}^{\xi_2+\xi_2^*}\nonumber\\
&&\quad \quad +
\left(\frac{c_{12}c_{21}-c_{11}c_{22}}{(p_2^*+p_1)(p_1^*+p_2)}
+\frac{c_{11}c_{22}-c_{12}c_{21}}{(p_1^*+p_1)(p_2^*+p_2)}
\right){\rm e}^{\xi_1+\xi_2+\xi_1^*+\xi_2^*}
\,,\\
&&g={\small\left|\matrix{
\displaystyle\frac{{\rm e}^{\xi_1+\xi_1^{*}}}{p_1+p_1^*} 
&\displaystyle\frac{{\rm e}^{\xi_1+\xi_2^{*}}}{p_1+p_2^*} 
&1&0
&{\rm e}^{\xi_1}
\cr\noalign{\vskip5pt}
\displaystyle\frac{{\rm e}^{\xi_2+\xi_1^{*}}}{p_2+p_1^*} 
&\displaystyle\frac{{\rm e}^{\xi_2+\xi_2^{*}}}{p_2+p_2^*} 
&0&1
&{\rm e}^{\xi_2}
\cr\noalign{\vskip5pt}
-1&0&
\displaystyle
\frac{\int_{-\infty}^{\infty}a_1^*a_1dy}{p_1^*+p_1}
&\displaystyle
\frac{\int_{-\infty}^{\infty}a_1^*a_2dy}{p_1^*+p_2}
&0
\cr\noalign{\vskip5pt}
0&-1&
\displaystyle
\frac{\int_{-\infty}^{\infty}a_2^*a_1dy}{p_2^*+p_1}
&\displaystyle
\frac{\int_{-\infty}^{\infty}a_2^*a_2dy}{p_2^*+p_2}
&0
\cr\noalign{\vskip5pt}
0&0&-a_1&-a_2&0
}\right|}\nonumber\\
&&\quad =a_1{\rm e}^{\xi_1}+a_2{\rm e}^{\xi_2}
+\frac{(c_{12}a_1-c_{11}a_2)(p_1-p_2)}
{(p_1^*+p_1)(p_1^*+p_2)}{\rm e}^{\xi_1+\xi_1^*+\xi_2}\\
&&\qquad \qquad \qquad  +\frac{(c_{22}a_1-c_{21}a_2)(p_1-p_2)}
{(p_2^*+p_1)(p_2^*+p_2)}{\rm e}^{\xi_2+\xi_2^*+\xi_1}\,,\\
&&g^*=-
{\small \left|\matrix{
\displaystyle\frac{{\rm e}^{\xi_1+\xi_1^{*}}}{p_1+p_1^*}
&\displaystyle\frac{{\rm e}^{\xi_1+\xi_2^{*}}}{p_1+p_2^*}
&1&0&0
\cr\noalign{\vskip5pt}
\displaystyle\frac{{\rm e}^{\xi_2+\xi_1^{*}}}{p_2+p_1^*}
&\displaystyle\frac{{\rm e}^{\xi_2+\xi_2^{*}}}{p_2+p_2^*}
&0&1&0
\cr\noalign{\vskip5pt}
-1&0&\displaystyle
\frac{\int_{-\infty}^{\infty}a_1^*a_1dy}{p_1^*+p_1}
&\displaystyle
\frac{\int_{-\infty}^{\infty}a_1^*a_2dy}{p_1^*+p_2}
&a_1^*
\cr\noalign{\vskip5pt}
0&-1&\displaystyle
\frac{\int_{-\infty}^{\infty}a_2^*a_1dy}{p_2^*+p_1}
&\displaystyle
\frac{\int_{-\infty}^{\infty}a_2^*a_2dy}{p_2^*+p_2}
&a_2^*
\cr\noalign{\vskip5pt}
-{\rm e}^{\xi_1^{*}}
&-{\rm e}^{\xi_2^{*}}
&0&0&0
}\right|}\nonumber\\
&&\quad =a_1^*{\rm e}^{\xi_1^*}+a_2^*{\rm e}^{\xi_2^*}
+\frac{(c_{21}a_1^*-c_{11}a_2^*)(p_1^*-p_2^*)}
{(p_1^*+p_1)(p_1+p_2^*)}{\rm e}^{\xi_1+\xi_1^*+\xi_2^*}\\
&&\qquad \qquad \qquad 
+\frac{(c_{22}a_1^*-c_{12}a_2^*)(p_1^*-p_2^*)}
{(p_2+p_1^*)(p_2+p_2^*)}{\rm e}^{\xi_2+\xi_2^*+\xi_1^*}\,, 
\end{eqnarray*}
where $c_{ij}=
\int_{-\infty}^{\infty}a_i^*a_jdy/(p_i^*+p_j)$. 

To make four localized pulses, i.e. 
$(2,2)$-localized pulse solution, 
we consider 
$a_i(y)=\sum_{j=1}^2 \alpha_{2(i-1)+j} 
{\rm sech}(k(y+\eta_j))$. Then $c_{ij}$ is given as follows. 
\begin{eqnarray*}
&&c_{ij}=\frac{(2/k)(\alpha_{2(i-1)+1}^*\alpha_{2(j-1)+1} 
+\alpha_{2(i-1)+2}^*\alpha_{2(j-1)+2})
}
{(p_i^*+p_j)}\\ 
&& \quad \quad +\frac{
4(\eta_1-\eta_2)(\alpha_{2(j-1)+1}\alpha_{2(i-1)+2}^*
+\alpha_{2(i-1)+1}^*\alpha_{2(j-1)+2}
)}
{({\rm e}^{k(\eta_1-\eta_2)}-{\rm e}^{-k(\eta_1-\eta_2)})
(p_i^*+p_j)} \,.
\end{eqnarray*}

\begin{figure}[t!]
\centerline{
\includegraphics[scale=0.7]{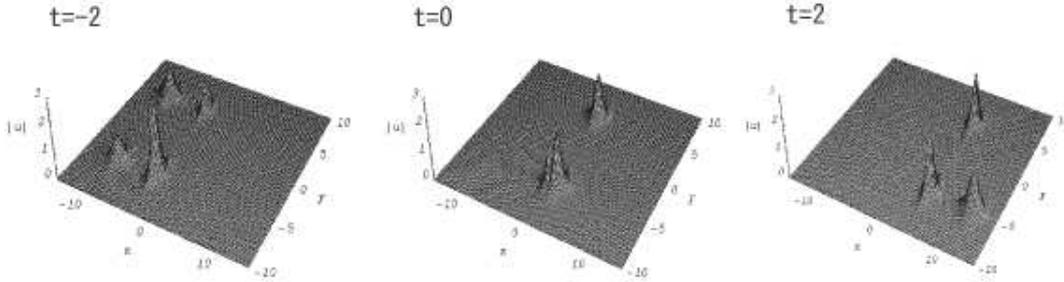}}
\caption{(2,2)-localized pulse solution. 
$\alpha_1=1+i, \alpha_2=1, \alpha_3=1, \alpha_4=1, 
p_1=3/2-5i/2, 
p_2=3-i,k=2,
\eta_1=-5, \eta_2=5$.}
\label{fig3}
\end{figure}

Figure \ref{fig3} is an example of $(2,2)$-localized pulse solution. 
It is observed that 4 localized pulses suddenly change the height of pulses
after a collision. Each pair of pulses on lines parallel to the $x$-axis 
collides, then the total mass of 
pulses is redistributed. In the case of figure \ref{fig3}, the height
of a localized pulse become very small after a collision. 
Although there is a distance between two pulses on a line parallel to the 
$x$-axis and other two pulses on another line, 
the collision causes an effect of 4-pulse interaction.
As this example, solutions of the 2DNNLS equation have very complicated
and interesting properties.  

\section{Conclusion}

We have discussed an integrable 2DNNLS equation and 
shown that the $N$-soliton solution of 
the 2DNNLS equation is given by the Gram type determinant and solutions
can be localized in $x$-$y$ plane. 

Note that the integrable 2DNNLS equation discussed in this Letter can be
considered as the vector NLS equation with infinitely many components
\cite{manakov,laksh,Miller,APT}. 
This fact suggests
that the vector soliton equations can produce nonlocal multi-dimensional
soliton equations having localized pulses. 

It should be noted that 
a model for second harmonic generation, i.e., quadratic solitons, 
was discussed in the paper by Nikolov et al.,
and they discussed the relationship between a nonlocal soliton equation 
and a vector soliton system \cite{Kroli2}. 
Finding physical systems which could be described by the 2DNNLS
equation is an interesting problem.  

{\bf Note added in proof:} 
After the acceptance of this Letter for publication, 
the authors noticed the 2DNNLS equation (2) is  equivalent to eq.(7.86) 
in ref.[11]. However, as far as we know, the N-soliton solution has not 
been obtained so far.  The authors thank Dr. Takayuki Tsuchida for
letting us know the paper by Zakharov [11].
 
 




\begin{thebibliography}{Oo}
\def\title#1{{\rm#1}}

\bibitem{APT}
M. J. Ablowitz, B. Prinari and A. D. Trubatch,  
{\it Discrete and Continuous Nonlinear Schr\"odinger Systems}
(Cambridge University Press, 2004).

\bibitem{pelinovsky}
D. Pelinovsky,  
Phys. Lett. A {\bf 197} (1995) 401.

\bibitem{pelinovsky2}
D. Pelinovsky and R. H. J. Grimshaw,   
J. Math. Phys. {\bf 36} (1995) 4203.

\bibitem{Kroli1}
W. Kr\'olikowski and O. Bang,  
Phys. Rev. E {\bf 63} (2000) 016610.

\bibitem{Kroli2}
N. I. Nikolov, D. Neshev, W. Kr\'olikowski and O. Bang,  
Phys. Rev. E {\bf 68} (2003) 036614.

\bibitem{Deco}
B. Deconinck and J. N. Kutz,  
Phys. Lett. A {\bf 319} (2003) 97.

\bibitem{HirotaBook} 
R. Hirota, {\it The Direct Method in Soliton Theory} 
(Cambridge University Press, 2004).

\bibitem{manakov}
S. V. Manakov,
Sov. Phys. JETP {\bf 38} (1974) 248.

\bibitem{laksh}
R. Radhakrishnan, M. Lakshmanan, and J. Hietarinta,  
Phys. Rev. E {\bf 56} (1997) 2213.

\bibitem{Miller}
P. D. Miller,  
Phys. Lett. A {\bf 101} (1997) 17.

\bibitem{zakharov}
V. E. Zakharov,  in Solitons, ed. Bullough and Caudrey, Topics in
	Current Physics (Springer, Berlin - New York, 1980) 243.

\end{thebibliography}
\end{document}